\documentclass[a4paper,11pt]{article}
\usepackage{jheppub}

\usepackage{multirow, graphicx,amssymb,url,mathrsfs,amsmath}
\usepackage{wrapfig,boxedminipage,setspace,subfigure,epsfig}
\usepackage{amsxtra,amstext,latexsym,dsfont,amsfonts}
\usepackage{color,eucal}
\usepackage[dvipsnames]{xcolor}
\usepackage{float}
\usepackage{slashed,comment}
\usepackage{tensor}
\usepackage{indentfirst}

\title{Pole-skipping of Holographic Correlators: Aspects of Gauge Symmetry and Generalizations}

\author{Yuan-Tai Wang}
\author{and Wen-Bin Pan}

\emailAdd{wangyuantai19@mails.ucas.ac.cn}

\emailAdd{panwenbin18@mails.ucas.ac.cn}
\affiliation{School of Physical Sciences, University of Chinese Academy of Sciences, Zhongguancun east road 80, Beijing 100190, China}

\abstract{In the framework of anti-de Sitter space/conformal field theory (AdS/CFT), we study the pole-skipping phenomenon of the holographic correlators of boundary operators. We explore the locations of the pole-skipping points case by case with the models of $U(1)$-gauged form fields propagating in the asymptotic AdS bulk of finite temperature. In general, in different cases all the first-order points are located at the Matsubara frequency with corresponding wave vectors regularly dispersed in the momentum space. Specifically, in the massless cases with $U(1)$ symmetry, the wave vectors of the pole-skipping points have a form-number dependence, and a trans-mode equivalence in the dual fields is found in correspondence with electromagnetic duality. In the massive cases with explicit symmetry breaking, we find that the appearance of a non-zero mass yields extra pole-skipping points which reduce to the massless results in zero mass limit. We expect in such kind of pole-skipping properties implications of distinctive physics in the chaotic systems. Our near-horizon computation is verified with the double-trace method especially in the example of 2-form where there is dimension-dependent boundary divergence. We illustrate in these cases that the pole-skipping properties of the holographic correlators are determined by the IR physics, consistent with the ordinary cases in previous studies.}

\keywords{AdS-CFT Correspondence, Gauge Symmetry, Renormalization and Regularization}

\arxivnumber{2209.04296}

\begin{document}
\maketitle
\flushbottom

\section{Introduction}\label{1}

It was originally conjectured by Maldacena in 1997 that the geometry of anti-de Sitter space (AdS) gravity in five dimensions has a physical equivalence with conformal field theory (CFT) living in four dimensions \cite{Maldacena:1997re}. High energy theorists recognized this correspondence as a holographic duality which connects gravitational theories as the long-range limit of string theory and strongly coupled quantum manybody systems described by the boundary CFT in one lower dimension \cite{Zaanen:2015oix, Ammon2015GaugeGravityDF, Hartnoll:2016apf}. This holographic principle is most famously exemplified by the gravitational theory living in the AdS bulk dual to a gauge field theory on the boundary, and it is therefore called by the high energy community AdS/CFT correspondence or gauge/gravity duality in the sense above \cite{Horowitz:2006ct, Hubeny:2014bla, Maldacena:2011ut, McGreevy:2009xe, Ramallo:2013bua}. Plenty of new theories and applications were put forward thereafter, bridging the gap between AdS as well as other generalized models of gravity and condensed matter physics as well as quantum information, known as AdS/CMT and AdS/QI respectively.

Starting off with GKPW relation \cite{Zaanen:2015oix, McGreevy:2009xe, Gubser:1998bc, Witten:1998qj}, one could make an identification of the partition functions on the two sides, which is reduced to $Z[J]\sim e^{-S_g}$ at large $N$ limit and t' Hooft limit $g^2_{YM}N \gg 1$, where $Z[J]$ is the generating functional of the CFT with a source $J$ and $S_g$ is the gravitational action. Then with the assumption of a bulk field $\phi$ as a perturbation on the background spacetime, which has a dual field operator $O$, the equality and the formulation of path integration together yield a substitutive formula for the correlation functions of the dual operators \cite{Zaanen:2015oix, McGreevy:2009xe, Gubser:1998bc, Witten:1998qj}: $G_n(x) = \left\langle O(x_1)\cdots O(x_n)\right\rangle = \frac{\delta^n S_g[\phi |_{\partial}=\phi_0]}{\delta\phi_{0}(x_1)\cdots\delta\phi_{0}(x_n)}|_{\phi_{0}=0}$, where $\phi$ takes the value of $\phi_0$ on the boundary. According to this holographic rationale, the computations in the field theory are successfully translated to the language of gravity.

Two decades later, studies of quantum many-body chaotic systems \cite{Blake:2018leo, Grozdanov:2017ajz, Blake:2017ris} demonstrated that the two-point functions of energy density display non-unique behavior around some special points in momentum space by using out-of-time ordered correlation functions (OTOCs). A brief physical picture of chaos is that classically it explains the macroscopic phenomena in the microscopic view, and naturally that quantum chaos is expected to play a similar role, deeply related to transport and hydrodynamics\cite{Blake:2018leo, Blake:2017ris, Gu:2016oyy, Patel:2016wdy, Grozdanov:2018atb, Lucas:2017ibu}. The non-uniqueness of the correlation functions was expected to be universal for maximally chaotic systems, which, in holography, can be regarded as finite-temperature systems with a gravitational dual coupled to matter \cite{Blake:2018leo, Blake:2017ris, Blake:2019otz, Natsuume:2019xcy}. Precisely speaking, the correlation functions were detected to be not uniquely defined in that a pole and a zero would intersect at those points, hence denominated as the ``pole-skipping'' phenomenon. 

Computations in gravity reincarnated such kind of pole-skipping with boundary operators chosen to be the scalar operator, $U(1)$ current and energy-momentum tensor, etc. \cite{Blake:2019otz, Natsuume:2019xcy, Kim:2021xdz, Ceplak:2021efc, Liu:2020yaf, Ahn:2020bks, Wang:2022mcq,Jeong:2021zhz}. To date, all results have confirmed the existence of these special points, while the exact locations of the points are diverse in the momentum space, e.g., the Matsubara frequencies for scalar are \cite{Blake:2019otz, Kim:2021xdz} $\omega_n = -2ni\pi T,\ n = 1,2,\cdots$ where $T$ is the temperature of the background spacetime. In addition, there has been evidence for the physical relation between pole-skipping and quantum chaos, transport, and hydrodynamics, e.g., pole-skipping in hyperbolic black holes in \cite{Ahn:2020bks}, and pole-skipping of the sound channel with spontaneous symmetry breaking in translation in \cite{Jeong:2021zhz}.

It would be natural and heuristic to probe the pole-skipping properties of generalized models of Maxwell $U(1)$ symmetry. For one thing, gauge symmetry is a type of unphysical symmetry, relating different representations of the same physical state in QFT \cite{Bain2008-BAIRHG-2, Struyve:2011nz}. Whereas gauge symmetry will be explicitly broken simply if a quadratic mass term is included in the Maxwell action. To restore the gauge symmetry, one has to consider auxiliary fields, which have been incorporated in the theories of spontaneous symmetry breaking, massive Yang-Mills model, and Higgs-Kibble model, etc. \cite{Slavnov:2005ip, Bettinelli:2007eu}. For another thing, as a focal point in holographic electromagnetic duality, there exist higher-form theories coupled to gravity which correspond to the same boundary CFT as the Maxwell bulk theory in certain dimensions if the boundary conditions are appropriately chosen \cite{Marolf:2006nd, DeWolfe:2020uzb}. Higher-form symmetry also plays a key role in the study of generalized global symmetry, cf. \cite{Grozdanov:2016tdf, Grozdanov:2018ewh, Hofman:2017vwr, Gaiotto:2014kfa}.

Meanwhile, analytic computation reveals that the 2-form bulk field has a dimension-dependent boundary divergence, which can be renormalized with the double-trace method \cite{Grozdanov:2018ewh, Hofman:2017vwr}. In particular, the 2-form bulk ansatz is solvable in $\mathrm{AdS}_{d+2}$ spacetime considering the $\mathrm{SO}_{d+1}$ symmetry, whereas the analytic solutioin is difficult to find in AdS black hole background spacetime \cite{Hofman:2017vwr}. Fortunately, the pole-skipping properties seem detectable in the latter case: it has long been believed that the pole-skipping points of the holographic correlators can be determined at least in the previous ordinary examples \cite{Blake:2019otz, Natsuume:2019xcy, Ceplak:2021efc}, even oftentimes without knowing the exact bulk solution. Then a subsequent question is how the redefined boundary behavior with the double-trace coupling would affect the pole-skipping properties in the divergent cases.

This study aims to provide glimpses into the pole-skipping phenomenon of the $U(1)$-related models. This paper is organized as follows. In section \ref{2}, we first quickly review the electromagnetic duality in holography and demonstrate the pole-skipping of the scalar correlator, i.e., the simplest model. In section \ref{3} to \ref{5}, we discuss the $U(1)$ models configured as 1-form, 2-form, and generic $p$-form with/without a field mass, respectively. In section \ref{6}, we further examine the pole-skipping properties by studying the role of the field mass and the form number, and explore the connection to the theory of electromagnetic duality, and make a summary of this paper.

\section{Preliminaries (review)}\label{2}

Before the discussion of $U(1)$ models, we first introduce the background spacetime geometry considered for this work, and briefly review the concept of electromagnetic duality and the pole-skipping phenomenon.

\subsection{Spacetime configuration}

We consider the asymptotic $\mathrm{AdS}_{d+2}$ black hole spacetime with a planar horizon. The geometry is a solution of Einstein's equation with a negative cosmological constant $\Lambda=-d(d+1)/2$, whose metric takes the form in Eddington-Finkelstein coordinates $(v,r,x,y,\cdots)$:
\begin{equation}\label{metric}
   \mathrm{d}s^2 = -r^2 f(r)\mathrm{d}v^2 + \mathrm{d}v\mathrm{d}r + \mathrm{d}r\mathrm{d}v + h(r)(\mathrm{d}x^2+\mathrm{d}y^2+\cdots),
\end{equation}  
where we have set the AdS radius $L=1$, following the convention of \cite{Blake:2019otz}. We assume the asymptotic behavior on the horizon $r=r_0$ and at infinity $r\rightarrow\infty$:
\begin{equation}\label{asympt}
\begin{split}
\begin{aligned}
    f(r_0)=0,\; f'(r_0)&=4 \pi T/r_0^2, \\
    f(r\rightarrow\infty)\sim 1,\; h(r\rightarrow\infty)&\sim r^2,\; Z(r\rightarrow\infty)\sim 1,
\end{aligned}
\end{split}
\end{equation}  
where $T$ denotes the black hole temperature, $Z(\phi(r))$, assumed to be a radial function, denotes the dilatonic coupling arising from dimensional reduction, and the prime sign denotes the derivative of $r$. We leave the bulk behavior of the functions unspecified, corresponding to different couplings of matter fields.

\subsection{Electromagnetic duality}\label{2.2}

Next, we present a physical picture of electromagnetic duality in asymptotic AdS, referring to \cite{DeWolfe:2020uzb}.

A generic $p$-form field $P$ propagating in $\mathrm{AdS}_{d+2}$ behaves on the boundary as \footnote{In \cite{DeWolfe:2020uzb}, the boundary spacetime is assumed to be $\mathrm{AdS}_{d+1}$. Please note this number difference from our convention for a close check.}
\begin{equation}\label{bdry_p_form}
    P(r\rightarrow\infty) = \alpha^{(p)} + \beta^{(p)} \; r^{2p-d-1} + \cdots,
\end{equation}
where we have chosen the gauge condition $P_{r\cdots}=0$. Both $\alpha$ and $\beta$ are constants of $r$, while $\alpha$ inherits the dynamical gauge symmetry $\alpha \rightarrow \alpha + \mathrm{d}\lambda$ and $\beta$ obeys the conservation law $\mathrm{d}\ast\beta=0$, where $\ast$ denotes the $(d+1)$-dimensional Hodge star. We carry out a more complete discussion on the theory of $p$-form in section \ref{5.1}, with the field components and spacetime coordinates specified.

The boundary solution is required to be fixed with some boundary condition. By convention, two types of boundary conditions are considered:
\begin{equation}\label{BC}
\begin{split}
\begin{aligned}
    &\alpha^{(p)} = \text{fixed},\quad \text{Dirichlet (regular) B.C.},\\
    \text{or}\;&\beta^{(p)} = \text{fixed},\quad \text{Neumann (alternate) B.C.},\\
\end{aligned}
\end{split}
\end{equation}  
or generically a mixed boundary condition. Given a certain boundary condition, the fixed term $\alpha (\text{or}\;\beta)$ is regarded as the source of the boundary current operator $J$ while the dynamical d.o.f. of the boundary CFT are contained in the other term $\beta (\text{or}\;\alpha)$, giving rise to the global (or dynamical) $U(1)$ symmetry for the field theory. According to the GKPW relation, one can check this argument with the substitutive formula imposed on the one-point function:
\begin{equation}
\begin{split}
\begin{aligned}
    \left\langle J^{(p)}\right\rangle &= \frac{\delta S_{tot}}{\delta\alpha^{(p)}} \propto \beta^{(p)},\quad \text{Dirichlet (regular) B.C.},\\
    \left\langle J^{(p)}\right\rangle &= \frac{\delta S_{tot}}{\delta\beta^{(p)}} \propto \alpha^{(p)},\quad \text{Neumann (alternate) B.C.},\\
\end{aligned}
\end{split}
\end{equation}
where a boundary counter term needs to be chosen appropriately and added to the total action $S_{tot}$.

The dual $(d-p)$-form field $Q$ behaves likewise on the boundary as
\begin{equation}
    Q(r\rightarrow\infty) = \Tilde{\alpha}^{(d-p)} + \Tilde{\beta}^{(d-p)} \; r^{d-2p-1} + \cdots,
\end{equation}
with a different set of constants $\Tilde{\alpha}$ and $\Tilde{\beta}$ at the first two leading orders. The boundary condition can be chosen in the same manner as eq. \eqref{BC}.

Electromagnetic duality relates the dual fields with a concise equality, i.e.,
\begin{equation}
    \mathrm{d}Q = \ast\; \mathrm{d} P.
\end{equation}
This relation yields a proportionality of the boundary terms of the dual fields, i.e.,
\begin{equation}\label{dual_prop}
    \Tilde{\beta}^{(d-p)} \propto \ast \mathrm{d}\alpha^{(p+1)},\quad \mathrm{d}\Tilde{\alpha}^{(d+1-p)} \propto \ast \beta^{(p)}.
\end{equation}
Therefore, the Dirichlet (regular) boundary condition for one field is translated to Neumann (alternative) boundary condition for its dual through electromagnetic duality. A less abstract example of 1-form/2-form duality in $d=3$ is also analyzed in \cite{DeWolfe:2020uzb}, for further reading.

\subsection{Pole-skipping of scalar}\label{2.3}

Now we come back to a minimally coupled free scalar, or 0-form, propagating in asymptotic AdS with the spacetime metric in the form of eq. \eqref{metric}. The scalar model has been studied extensively in various literature as a typical example of the gravitational theory coupled with matter, and hereby we only present a minimal content of the story with an emphasis on finding the pole-skipping points, following \cite{Blake:2019otz}.

The bulk action with a massive scalar field $\varphi$ is
\begin{equation}
    S[g,\varphi]=\int\mathrm{d}^{d+2} x \sqrt{-g}\big(R-2\Lambda-\frac{1}{2}(\partial^2\varphi+m^2\varphi^2)\big).
\end{equation}
By taking the functional derivatives of the field, one obtains EOM \footnote{In this paper, we use Greek letters for bulk indices, and will use Latin letters for boundary indices or some bulk cases when the radial index $r$ is ruled out.}
\begin{equation}
    (\nabla^2-m^2)\varphi = \big(\frac{1}{\sqrt{-g}}\partial_M(\sqrt{-g}\partial^M)-m^2\big)\varphi = 0.
\end{equation}
We assume that the wave vector $k$ of the field lies in $x$ direction considering the rotational spatial symmetry, and that the field has a plane-wave expansion
\begin{equation}
    \varphi(r,v,x)=\int \mathrm{d}^2k \; e^{-i\omega v+ikx}\phi(r,\omega,k).
\end{equation}
Then in momentum space the EOM reduces to
\begin{equation}\label{EOM_scalar}
    \big( h^{d/2}(r^2f\phi'-i\omega\phi) \big)' - i\omega h^{d/2}\phi' - h^{d/2-1}(k^2+m^2h)\phi = 0,
\end{equation}
where the prime sign denotes the derivative of $r$. The boundary solution is at leading order
\begin{equation}\label{bdry_scalar}
    \phi(r\rightarrow\infty) = \phi_A(\omega,k)r^{\Delta-d-1} + \phi_B(\omega,k)r^{-\Delta},
\end{equation}
where $\Delta$ is the larger root of $\Delta(\Delta-d-1)=m^2$, which is also the conformal dimensioin of the dual operator $O$. Therefore, with the standard procedure, the (retarded) holographic correlator of dual operators is evaluated as
\begin{equation}\label{GF_scalar}
     G^R_{O,O}(\omega,k) \propto \frac{\phi_B(\omega,k)}{\phi_A(\omega,k)}.
\end{equation}
Requiring the asymptotic behavior in eq. \eqref{asympt} and imposing the ansatz of field expansion near the horizon
\begin{equation}\label{horizon_ansatz_scalar}
     \phi(r)=(r-r_0)^{\lambda}(\phi^{(0)}+\phi^{(1)}(r-r_0)+\phi^{(2)}(r-r_0)^2+\cdots),
\end{equation}
one can expand eq. \eqref{EOM_scalar} in powers of $r-r_0$ obeying the in-falling boundary condition
\begin{equation}
    \eqref{EOM_scalar}\equiv \mathcal{S} = \mathcal{S}_0 + \mathcal{S}_1 (r-r_0) + \mathcal{S}_2 (r-r_0)^2 + \cdots = 0,
\end{equation}
and solve the equations of the coefficients order by order. The lowest order equation is simply
\begin{equation}
\begin{split}
\begin{aligned}
    &\lambda (i\omega-2\pi T\lambda) = 0, \\
    \Longrightarrow  &\lambda_1 = 0,\quad \lambda_2 = \frac{i\omega}{2\pi T}. \\
\end{aligned}
\end{split}
\end{equation}

Obviously, with generic $\omega$, $\lambda_1$ corresponds to a unique in-falling solution and $\lambda_2$ corresponds to an out-going solution. The two solutions degenerate when $\omega=-2ni\pi T,\;n=1,2,\cdots$, rendering the field expansion non-unique. For a close check, one would find the second-lowest order equation with $\lambda=\lambda_1$ to be
\begin{equation}\label{first_order_eq.}
    M_{11}\phi^{(0)} + (4\pi T-2i\omega)h\phi^{(1)} = 0,
\end{equation}
where $h$ is evaluated on the horizon $r=r_0$ and $M_{11} = -(k^2+m^2h+i\omega dh'/2)$. Generically, $\phi^{(1)}$ can be uniquely determined with $\phi^{(0)}$ set free, and likewise higher-order coefficients can be determined by iteration. The equation becomes singular when $\omega = -2i\pi T$, with the second term vanishing. This is fine, however, when $M_{11} \neq 0$ and therefore $\phi^{(0)}$ must vanish, leaving $\phi^{(1)}$ to be a free parameter and higher-order coefficients can be fixed thereon. Nonetheless, both $\phi^{(0)}$ and $\phi^{(1)}$ are free when $M_{11} = 0$ and consequently the undetermined solution has extra d.o.f.. Therefore, the first-order special points are claimed to be found at
\begin{equation}\label{ps_massive_scalar}
    \omega = -2i\pi T,\quad M_{11} = 0 \;\Rightarrow\; k^2 = -m^2 h - \pi Tdh', \\
\end{equation}
where $h$ is evaluated at $r=r_0$. Higher-order special points can be found by solving $S_n = 0$ iteratively. A nice trick is to rewrite these equations into a matrix form $M\cdot\Phi = 0$ with $\Phi = [\phi^{(0)},\;\phi^{(1)},\;\phi^{(2)},\cdots]^T$ and corresponding coefficients in $M$. One can check that the conditions $M_{n,n+1}=0,\;\mathrm{det}M^{[n]}=0$ yield the special points at $\omega=-2ni\pi T,\;n=1,2,\cdots$ and some $2n$ values of $k$, where $M^{[n]}$ denotes the first $n$ rows and columns of $M$.

At these points, a pole and a zero would intersect: slightly move away from the special point to $\omega\rightarrow\omega+\epsilon\delta\omega,\;k\rightarrow k+\epsilon\delta k$ and one would find from eq. \eqref{first_order_eq.} that the coefficients of the series expansion are related, with a dependence on the slope $\delta\omega/\delta k$. We interpret these special points found with near-horizon analysis as the pole-skipping points of the boundary correlator. Such an interpretation makes sense in that there is no boundary divergence for the bulk scalar, cf. eq. \eqref{bdry_scalar}, and therefore unambiguously, the correlator can be obtained from the on-shell solution, cf. eq. \eqref{GF_scalar}, which depend on the near-horizon boundary condition and consequently on the slope $\delta\omega/\delta k$.

With all the discussion above, we are well-prepared to move on to the $U(1)$ models in the following sections.

\section{1-form}\label{3}
To begin with, we evaluate the pole-skipping points of 1-form gauge currents by near-horizon analysis. In subsection \ref{3.1}, we briefly review the analytic results of the longitudinal mode in \cite{Blake:2019otz, Natsuume:2019xcy}, and the transverse mode in \cite{Natsuume:2019xcy}. In subsection \ref{3.2}, we generalize the dual bulk field with a non-zero mass and present two equivalent methods of finding the corresponding pole-skipping points, referring to \cite{Jimenez-Alba:2014iia, Jimenez-Alba:2015awa}.

\subsection{Massless 1-form (review)}\label{3.1}

The action term of a minimally-coupled 1-form field $A$ is \cite{Blake:2019otz}
\begin{equation}
    S[A]=-\frac{1}{4}\int\mathrm{d}^{d+2} x \sqrt{-g}Z(\phi)F^2,
\end{equation}
where $Z(\phi)$ is the dilatonic coupling, and $F=\mathrm{d}A$ is the field strength, with components $F_{MN}=\partial_{M}A_{N}-\partial_{N}A_{M}$. For simplicity, we will assume the dilaton field $\phi$ to be non-dynamical, i.e., dependent only on the radial coordinate $r$ considering the static background spacetime. The theory has $U(1)$ gauge symmetry, i.e., $A_{M} \rightarrow A_{M} + \partial_M \Lambda,\ \forall \Lambda$. By taking the functional derivatives, one obtains EOM
\begin{equation}
    \partial_M(\sqrt{-g}Z(\phi)F^{MN}) = 0.
\end{equation}

We still assume that the wave vector $k$ of the field lies in $x$ direction and that the field has a plane-wave expansion, i.e., 
\begin{equation}
    A(r,v,x)=\int \mathrm{d}^2k \; e^{-i\omega v+ikx}a(r,\omega,k).
\end{equation}
We fix the gauge symmetry with condition $A_r=0$. With the same asymptotic AdS background spacetime setting as in eq. \eqref{metric}, we find a dimension-dependent mode decomposition of 1-form components: for $d=2$, the parity transformation $y\rightarrow -y$ yields the longitudinal mode containing $a_v, a_x$ and the transverse mode containing $a_y$; for $d\geq 3$, the spatial $SO(d-1)$ symmetry and the corresponding transformation rule yield the longitudinal mode containing $a_v,\ a_x$ and the transverse mode containing $a_{\mathbf{z}}$ where $\mathbf{z}$ denotes an arbitrary spatial coordinate except $x$. Therefore, the results of decomposition in different spacetime dimensions are highly consistent, although originating from different kinds of symmetry. For simplicity, we call the two modes to be longitudinal mode and transverse mode respectively. EOM are simply evaluated as \footnote{To derive EOM in the form of field components from the covariant equation, the inverse metric must be used to contract the covariant(lower-script) field components, since forms are defined as anti-symmetric covariant tensors.}
\begin{equation}\label{EOM_massless_1_form}
\begin{split}
\begin{aligned}
    0 &= k^2 a_v + k\omega a_x + i\omega h {a_v}' + ik r^2 f {a_x}', \\
    0 &= (\frac{1}{2}dZh'+hZ'){a_v}' + ikZ {a_x}' + hZ {a_v}'',\\
    0 &= (Z(2k^2+i(d-2)\omega h')+2i\omega hZ')a_{\mathbf{z}} - 2r^2 fhZ a_{\mathbf{z}}''\\
      &+ ((2-d)r^2 fZh'-2h(Z(-2i\omega+2rf+r^2 f')+r^2 fZ'))a_{\mathbf{z}}'.\\
\end{aligned}
\end{split}
\end{equation}
Requiring the asymptotic behavior in eq. \eqref{asympt} and imposing the ansatz of field expansion
\begin{equation}\label{horizon_ansatz_1_form}
     a_{\mu}(r)=(r-r_0)^{\lambda_{\mu}}(a_{\mu}^{(0)}+a_{\mu}^{(1)}(r-r_0)+a_{\mu}^{(2)}(r-r_0)^2+\cdots),
\end{equation}
the decoupled EOM for the transverse mode can be analysed following the order-by-order computation of the scalar model in section \ref{2.3}. As for the longitudinal mode, one could either decouple the EOM or solve the set by expanding all the field components, or transform the order-by-order equations into matrix equations and solve them as an eigenvalue problem \cite{Natsuume:2019xcy}. No matter what trick one employs, the gauge redundancy, distinct from the example of scalar field, would leave $a_{\mu}^{(0)}$ undetermined and the higher-order coefficients shall depend on $a_{\mu}^{(1)}$ by iteration. All in all, one can find the first-order special points for the two modes at
\begin{equation}\label{ps_massless_1_form}
\begin{split}
\begin{aligned}
    \omega&=-2i\pi T,\ \ k^2=\pi T((d-2)h'+2hZ'/Z),\quad \text{for longitudinal mode},\\
    \omega&=-2i\pi T,\ \ k^2=-\pi T((d-2)h'+2hZ'/Z),\ \text{for transverse mode},\\
\end{aligned}
\end{split}
\end{equation}
where $h$ and $Z$ are evaluated at $r=r_0$. 

Caveat: strictly speaking, the longitudinal mode has a zeroth order pole-skipping point at $\omega=k=0$ while the transverse mode does not. One can check this subtlety by first setting $k=0$ in the first two equations of eq. \eqref{EOM_massless_1_form} and then making a classified discussion on the value of $\omega$. That is, the leading-order point is mode-dependent. These results are consistent with \cite{Blake:2019otz, Ahn:2020bks}. Such kind of pole-skipping at the origin takes place in the cases of higher forms as well, while $k$ may be shifted to some non-zero values in other configurations, e.g., \cite{Wang:2022mcq}. We will only focus on the non-zero points in the following sections concerning the fact that the $(0,0)$ point corresponds to the hydrodynamic limit where the behavior of the holographic correlators can be probed by hydrodynamics \cite{Blake:2019otz}.

One can check that similar to the case of the scalar field, there is no divergence in the boundary solutions of 1-form. Therefore, these special points found with near-horizon analysis are likewise the pole-skipping points of the boundary correlators of dual currents. The same is true of the massive 1-form in the next subsection. One can also check that as an alternative method, the special points of gauge invariants are the same as those we have found for the longitudinal mode. Note that in the viewpoint of the boundary CFT, the correlators of the same mode are related by conformal Ward identity \cite{Blake:2019otz, Corley:2000ev} and have the same set of pole-skipping points as a consequence.

\subsection{Massive 1-form}\label{3.2}

The 1-form action added with a quadratic term of a non-zero mass $m$ is
\begin{equation}
    S[A]=-\frac{1}{4}\int\mathrm{d}^{d+2} x \sqrt{-g}Z(\phi)(F^2+2 m^2 A^2).
\end{equation}
The corresponding EOM are
\begin{equation}\label{EOM_massive_1_form}
    \partial_M(\sqrt{-g}Z(\phi)F^{MN})-\sqrt{-g}Z(\phi)m^2 A^N = 0.
\end{equation}
The mass in the action explicitly breaks the gauge symmetry. The full set of EOM, containing a large number of coupled terms, is too complicated to solve. A shortcut is to perform the derivative operator $\partial_N$ on EOM, which yields a constraint equation
\begin{equation}
    \partial_M(\sqrt{-g}Z(\phi) A^M) = 0.
\end{equation}
We hereby point out that this constraint holds only on-shell and does not imply global $U(1)$ symmetry. We rewrite this equation as
\begin{equation}
\begin{split}
\begin{aligned}
    \partial_M A^M &= -\frac{1}{\sqrt{-g}Z(\phi)} A^M \partial_M(\sqrt{-g}Z(\phi))\\
    &= -\frac{1}{\sqrt{-g}Z(\phi)} A^r \partial_r(\sqrt{-g}Z(\phi)).
\end{aligned}
\end{split}
\end{equation}
Inserting this constraint into eq. \eqref{EOM_massive_1_form}, we obtain EOM of a simplified version: \footnote{The EOM in the form of field components are lengthy, which we omit here. $a_r$, $a_v$, $a_x$ components are coupled in our configuration.}
\begin{equation}
\begin{split}
\begin{aligned}
    0 &= 
    \partial_M(\sqrt{-g}Z(\phi)\partial^M A^N) - \partial_M\big(\sqrt{-g}Z(\phi)(g^{MP}\partial_P g^{NQ}-g^{NP}\partial_P g^{MQ})A_Q \big) \\
    &+ \sqrt{-g}Z(\phi)g^{NP}\partial_P\big(\frac{1}{\sqrt{-g}Z(\phi)}A^r\partial_r(\sqrt{-g}Z(\phi))\big) - \partial_r\big( \sqrt{-g}Z(\phi)g^{NP}\big)(\partial_P A^r) \\
    &- \sqrt{-g}Z(\phi)m^2 A^N.
\end{aligned}
\end{split}
\end{equation}

With the same ansatz of near-horizon expansion as eq. \eqref{horizon_ansatz_1_form}, we find the first pole-skipping points for the longitudinal mode at
\begin{equation}\label{ps_massive_1_form}
\begin{split}
\begin{aligned}
    \omega &= -2i\pi T,\\
    k^2 &= -\left( m^2 h(r_0)+ \pi T h'(r_0) \pm \pi T\sqrt{\left[(d-1)h'(r_0)+2h(r_0) \frac{Z'(r_0)}{Z(r_0)}\right]^2+\frac{4m^2}{\pi T}h(r_0)h'(r_0)}\right).\\
\end{aligned}
\end{split}
\end{equation}
In $m\rightarrow 0$ limit, one of the ($k^2$) reduces to the special points of the massless scalar, and the other reduces to the special points of the longitudinal mode of the massless 1-form, cf. eq. \eqref{ps_massless_1_form}. The appearance of a non-zero mass breaks the gauge symmetry and yields extra pole-skipping points.

Caveat: strictly speaking, the longitudinal mode has a zeroth order pole-skipping point at $\omega=0,\;k=-m^2 h(r_0)$. In $m\rightarrow 0$ limit, this point reduces to ($\omega$=k=0), i.e. the zeroth order point of the longitudinal mode of the massless 1-form. 

An equivalent method is to employ St\"{u}ckelberg mechanism. We briefly present the process as follows. With an auxiliary scalar field $\theta$, a gauge invariant action can be written as \cite{Jimenez-Alba:2014iia}
\begin{equation}
    S[A]=-\frac{1}{4}\int\mathrm{d}^{d+2} x \sqrt{-g}Z(\phi)(F^2+2 m^2 (A+\frac{1}{m}\mathrm{d}\theta)^2).
\end{equation}
The restored gauge symmetry is manifest: $A_{M} \rightarrow A_{M} + \partial_M \Lambda,\ \theta \rightarrow \theta -m \Lambda,\ \forall \Lambda$. With gauge invariants $F_{MN}$ and $J_M=A_M +\frac{1}{m}\partial_M\theta$, EOM are
\begin{equation}
\begin{split}
\begin{aligned}
    \nabla_\mu (Z(r) F^{\mu \nu})&=m^2 Z(r) J^\nu,\\
    \partial_{[\rho} F_{\mu \nu]}&=0,\\
    \nabla_\mu (Z(r) J^{\mu })&=0,\\
    \partial_{[\mu} J_{\nu]}&=F_{\mu \nu}.
\end{aligned}
\end{split}
\end{equation}
With the near-horizon expansion similar to eq. \eqref{horizon_ansatz_1_form} for the gauge invariants, one can find the first pole-skipping points for the longitudinal mode identical to the ones in eq. \eqref{ps_massive_1_form}.

\section{2-form}\label{4}

In this section, we study the pole-skipping phenomenon of holographic correlators of $U(1)$ currents generalized as 2-form. Specifically, we attempt to renormalize the UV divergence of 2-form following \cite{Grozdanov:2018ewh, Hofman:2017vwr}, and to calculate the redefined correlators. In subsection \ref{4.1}, we focus on the $U(1)$ gauge field; in subsection \ref{4.2}, we generalize the field configuration with a non-zero mass.

\subsection{Massless 2-form}\label{4.1}

The action term of a minimally-coupled 2-form field $B$ is \footnote{We have additionally include the dilatonic coupling in the action in consistency with our previous configuration.} \cite{Hofman:2017vwr}
\begin{equation}\label{action_massless_2_form}
    S=\frac{1}{6\gamma^2}\int \mathrm{d}^{d+2}x \sqrt{-g} Z(\phi) H^2,
\end{equation}
where we have made explicit the coupling $\gamma$, and $H=\mathrm{d}B$ is the field strength, with components $H_{MNP}=\partial_M B_{NP}+\partial_N B_{PM}+\partial_P B_{MN}$. The theory has $U(1)$ gauge symmetry, i.e., $B_{MN} \rightarrow B_{MN} + \partial_M \Lambda_N - \partial_N \Lambda_M,\ \forall \Lambda$. By taking the functional derivatives, one obtains EOM
\begin{equation}\label{EOM_massless_2_form}
    \partial_M(\sqrt{-g}Z(\phi)H^{MNP}) = 0.
\end{equation}
Then, the boundary current operator $J$ derived from on-shell action is
\begin{equation}\label{current_massless_2_form}
    J^{\mu\nu} = -\frac{1}{\gamma^2} \sqrt{-g}Z(\phi) H^{r\mu\nu}.
\end{equation}
$J$ has a continuation in the bulk when necessary.
We still assume that the wave vector of the field lies in $x$ direction, and the field has a plane-wave expansion, i.e., 
\begin{equation}
    B(r,v,x)=\int \mathrm{d}^2k \; e^{-i\omega v+ikx}b(r,\omega,k).
\end{equation}
We impose the gauge condition $B_{r\mu}=0$. Then in asymptotic AdS background spacetime, the mode decomposition is as follows: 
for $d\geq 3$, the spatial $SO(d-1)$ symmetry and the corresponding transformation rule yield the null mode containing $b_{vx}$ which is trivial, the longitudinal mode containing $b_{v\mathbf{z}}$ and $b_{x\mathbf{z}}$, and the transverse mode containing $b_{\mathbf{z_1}\mathbf{z_2}}$, where $\mathbf{z}$ denotes an arbitrary spatial coordinate except $x$. 

EOM of the longitudinal and transverse modes are evaluated as
\begin{equation}\label{physical_EOM_massless_2_form}
\begin{split}
\begin{aligned}
    0 &= k^2 b_{v\mathbf{z}} + k\omega b_{x\mathbf{z}} + i\omega h b_{v\mathbf{z}}' + ik r^2 f b_{x\mathbf{z}}',\\
    0 &= ikZ b_{x\mathbf{z}}' + ((d/2-1)Z h'+h Z')b_{v\mathbf{z}}' + hZ b_{v\mathbf{z}}'',\\
    0 &= (Z(2k^2+i(d-4)\omega h')+2i\omega hZ')b_{\mathbf{z_1}\mathbf{z_2}} - 2r^2 fhZ{b_{\mathbf{z_1}\mathbf{z_2}}}''\\
      &+ ((4-d)r^2 fZh'-2h(Z(-2i\omega+2rf+r^2 f')+r^2 fZ')){b_{\mathbf{z_1}\mathbf{z_2}}}'.\\
\end{aligned}
\end{split}
\end{equation}
This gauge redundancy is obviously seen in the longitudinal mode but harmless if one seeks to find the pole-skipping points, in that the coefficients $b^{(n)}$ of the near-horizon power series $b_{vec}(r)=b^{(0)}+b^{(1)} (r-r_0)+\cdots$ can be obtained by iteration from $b^{(1)}$ while $b^{(0)}$ can be left arbitrary. But this indeterminism in one single component, in the sense of UV, will obscure the definition of the correlator of boundary currents which depends on the near-horizon boundary condition exactly. As a remedy, we construct a gauge invariant $\beta_1$ and its orthogonal pair $\beta_2$:
\begin{equation}\label{transform_rule_massless_2_form}
    \beta_{1} = k b_{v\mathbf{z}} + \omega b_{x\mathbf{z}},\ \ \beta_{2} = \omega b_{v\mathbf{z}} - k b_{x\mathbf{z}},
\end{equation}
which yield EOM
\begin{equation}\label{gauge_invar_EOM_massless_2_form}
\begin{split}
\begin{aligned}
    0 &= (kZ(2i\omega+(d-2)h')+2khZ')\beta_{1}'+(Z(-2ik^2+(d-2)\omega h')+2\omega hZ')\beta_{2}'\\
      &+2khZ\beta_{1}''+2\omega hZ\beta_{2}'',
    \\
    0 &= k(k^2+\omega^2)\beta_{1} + ik\omega(r^2 f+h)\beta_{1}' + i(\omega^2 h-k^2 r^2 f)\beta_{2}'.
    \\
\end{aligned}
\end{split}
\end{equation}
Now $\beta_{1}$ is saved from the extra d.o.f., which are, not of our interest, fully described instead by an arbitrary constant term in $\beta_{2}$. Imposing the ansatz of near-horizon expansion for $b_{\mu\nu}$
\begin{equation}\label{horizon_ansatz_2_form}
     b_{\mu\nu}(r)=(r-r_0)^{\lambda_{\mu\nu}}(b_{\mu\nu}^{(0)}+b_{\mu\nu}^{(1)}(r-r_0)+b_{\mu\nu}^{(2)}(r-r_0)^2+\cdots),
\end{equation}
or equivalently for the gauge invariants, \footnote{For simplicity, we refer to the gauge invariant $\beta_{1}$ and its pair $\beta_{2}$ as gauge invariants hereinafter.} we find the first-order special points for the longitudinal and transverse modes at
\begin{equation}\label{ps_massless_2_form}
\begin{split}
\begin{aligned}
    \omega&=-2i\pi T,\ \ k^2=\pi T((d-4)h'+2hZ'/Z),\quad \text{for longitudinal mode},\\
    \omega&=-2i\pi T,\ \ k^2=-\pi T((d-4)h'+2hZ'/Z),\ \text{for transverse mode},\\
\end{aligned}
\end{split}
\end{equation}
where $h$ and $Z$ are evaluated at $r=r_0$. Moreover, similar to the case of 1-form, The 2-form longitudinal mode has an extra zeroth order pole-skipping point at $\omega=k=0$.

As we have argued, for scalar and 1-form fields, such kind of special points near the horizon are naturally regarded as the pole-skipping points of their boundary correlators. Now We demonstrate with the model of 2-form in the following that the same is true of higher-form fields which could be divergent on the boundary. Put it another way, pole-skipping, a property of the holographic correlators, is only determined by IR physics in the bulk spacetime, which would become self-evident once the boundary divergence is properly renormalized. 

Unlike the scalar and 1-form, the leading-order boundary behavior of 2-form depends on the spacetime dimension. Without loss of generality, we examine the boundary behavior and the boundary correlators of dual currents of the longitudinal mode in asymptotic $\mathrm{AdS}_{4}$ and $\mathrm{AdS}_{5}$. For both of the two cases, following \cite{Erdmenger:2015qqa}, we adopt the ansatz of boundary expansion for the gauge invariants
\begin{equation}\label{bdry_ansatz_2_form}
\begin{split}
\begin{aligned}
    \beta_{1}(r\rightarrow\infty)&=r^{\lambda_1}(b_{00}+\frac{b_{10}+b_{11}\mathrm{log}(r)}{r}+\cdots),\\
    \beta_{2}(r\rightarrow\infty)&=r^{\lambda_2}(b'_{00}+\frac{b'_{10}+b'_{11}\mathrm{log}(r)}{r}+\cdots),\\
\end{aligned}
\end{split}
\end{equation}
as well as the asymptotic behavior at infinity introduced in eq. \eqref{asympt}.

\begin{itemize}
    \item \bf{\Large{Asymptotic $\mathrm{AdS}_{4}$}}
\end{itemize}

In asymptotic $\mathrm{AdS}_{4}$ black hole (studied earlier in \cite{Grozdanov:2018ewh}), the boundary solution to the gauge-invariant EOM in eq. \eqref{gauge_invar_EOM_massless_2_form} is
\begin{equation}\label{bdry_solution_massless_2_form_4d}
\begin{split}
\begin{aligned}
    \beta_{1}(r\rightarrow\infty)&= jr,\\
    \beta_{2}(r\rightarrow\infty)&= \frac{2k\omega}{k^2-\omega^2}\ jr + const,\\
\end{aligned}
\end{split}
\end{equation}
where $j$ is the coefficient which will be fixed by the value of boundary current $J^1$ dual to the source $\beta_{1}(r\rightarrow\infty)$, and we have only kept the terms up to the first sub-leading order.
From eq. \eqref{current_massless_2_form}, the boundary currents $J^{\mu\nu}$ dual to $b_{\mu\nu}$ are simply
\begin{equation}\label{current_physical_massless_2_form_4d}
\begin{split}
\begin{aligned}
    J^{v\mathbf{z}} &= \frac{b_{v\mathbf{z}}'}{\gamma^2},
    \\
    J^{x\mathbf{z}} &= \frac{i(k b_{v\mathbf{z}}+\omega b_{x\mathbf{z}}+i r^2 b_{x\mathbf{z}}')}{r^2\gamma^2}.\\
\end{aligned}
\end{split}
\end{equation}
Therefore $J^1$ is at leading order
\begin{equation}\label{current_gauge_invar_massless_2_form_4d}
\begin{split}
\begin{aligned}
    J^{1}&=J^{\mu\nu}\frac{\delta b_{\mu\nu}}{\delta\beta_{1}}\\
    &\sim\frac{j}{\gamma^2(k^2-\omega^2)}=\frac{\beta_{1}'}{\gamma^2(k^2-\omega^2)},\\
\end{aligned}
\end{split}
\end{equation}
where in the second line we have inserted the transformation rule of the field components in eq. \eqref{transform_rule_massless_2_form} and the boundary solution in eq. \eqref{bdry_solution_massless_2_form_4d}. With $J^1$ we rewrite the boundary solution in eq. \eqref{bdry_solution_massless_2_form_4d} as
\begin{equation}\label{bdry_solution_massless_2_form_4d(1)}
    \beta_{1}(r\rightarrow\infty) = \gamma^2(k^2-\omega^2)J^1 r.
\end{equation}
The divergence as $r\rightarrow\infty$ makes the boundary behavior not well-defined, dependent on a radial cutoff $r_{\Lambda}$ at some energy scale $\Lambda$. Regarding the boundary CFT as a matrix-valued field theory, we consider $J$ as a single-trace operator \cite{Witten:2001ua}. Then with the double-trace deformation \cite{Witten:2001ua}
\begin{equation}\label{double_trace_deform}
    S_{ct} = \frac{\gamma^2}{2\kappa}\int\mathrm{d}^{d+1}x \; J_{\mu\nu}J^{\mu\nu},
\end{equation}
we obtain a shift term for the boundary solution, i.e.,
\begin{equation}\label{shift_massless_2_form_4d}
\begin{split}
\begin{aligned}
    \frac{\delta S_{ct}}{\delta J^1}&= \frac{\gamma^2}{2\kappa}\frac{\delta J_{\mu\nu}J^{\mu\nu}}{\delta J^1}\\
    &=\frac{\gamma^2}{2\kappa}\eta_{\mu\rho}\eta_{\nu\sigma}(J^{\rho\sigma}\frac{\delta J^{\mu\nu}}{\delta J^1}+J^{\mu\nu}\frac{\delta J^{\rho\sigma}}{\delta J^1})\\
     &=\frac{\gamma^2}{2\kappa}\eta_{\mu\rho}\eta_{\nu\sigma}(J^{\rho\sigma}\frac{\delta \beta_1}{\delta b_{\mu\nu}}+J^{\mu\nu}\frac{\delta \beta_1}{\delta b_{\rho\sigma}})\\
     &\sim -\frac{j}{\kappa},\\
\end{aligned}
\end{split}
\end{equation}
where in order to compute the leading-order term in the last line, we have used the Lorentzian metric $\eta_{\mu\nu}=\mathrm{diag}(-1,1,1)$ for the boundary field, the values of the boundary currents in eq. \eqref{current_physical_massless_2_form_4d}, and the transformation rule of the field components in eq. \eqref{transform_rule_massless_2_form}. At some cutoff $r_{\Lambda}$, this shift term and the boundary solution in eq. \eqref{bdry_solution_massless_2_form_4d(1)} yield altogether
\begin{equation}\label{bdry_solution_massless_2_form_4d(2)}
\begin{split}
\begin{aligned}
     \beta_1(r_{\Lambda})&=\gamma^2(k^2-\omega^2)J^1 r_{\Lambda} - \frac{\gamma^2(k^2-\omega^2)J^1}{\kappa}\\
     &\equiv b_{00} - \frac{\gamma^2(k^2-\omega^2)J^1}{\kappa},\\
\end{aligned}
\end{split}
\end{equation}
where in the second line we have defined $b_{00}$, which is finite now, as the source. Compare with the choices of boundary conditions in section \ref{2.2} and we know that eq. \eqref{bdry_solution_massless_2_form_4d(2)} corresponds to a mixed boundary condition.
Finally, we evaluate the correlator of $J^1$ in the linear response model:
\begin{equation}\label{GF_massless_2_form_4d}
\begin{split}
\begin{aligned}
     G^R_{J^1,J^1}(\omega,k) &= \frac{J^1}{b_{00}}\\
     &=\frac{J^1}{\beta_1(r_{\Lambda})+\frac{\gamma^2(k^2-\omega^2)J^1}{\kappa}}\\
     &=\frac{1}{\gamma^2(k^2-\omega^2)[\frac{\beta_1(r_{\Lambda})}{\beta_1'(r_{\Lambda})}+\frac{1}{\kappa}]}.\\
\end{aligned}
\end{split}
\end{equation}

The correlator, a physical quantity, should be independent of the cutoff scale \cite{Grozdanov:2018ewh}. By assuming the double-trace coupling $\kappa$ as a function of energy scale $\Lambda$, one can obtain from the square bracket in eq. \eqref{GF_massless_2_form_4d} a $\beta$-function and evaluate the fixed point by extremization.
Furthermore, since the bulk field EOM has no global solution, $\beta_1(r_{\Lambda})$ has to be expanded as the infinite series from the near-horizon expansion similar to eq. \eqref{horizon_ansatz_2_form}. Therefore, $G^R_{J^1,J^1}$ has a series of pole-skipping points, the first of which is exactly $(\omega,k)$ in eq. \eqref{ps_massless_2_form}, unaltered by the shift term resulting from the double-trace deformation. 

\begin{itemize}
    \item \bf{\Large{Asymptotic $\mathrm{AdS}_{5}$}}
\end{itemize}

In analogy, in asymptotic $\mathrm{AdS}_{5}$ black hole (studied earlier in \cite{Hofman:2017vwr}), the boundary solution to the gauge-invariant EOM in eq. \eqref{gauge_invar_EOM_massless_2_form} is
\begin{equation}\label{bdry_solution_massless_2_form_5d}
\begin{split}
\begin{aligned}
    \beta_{1}(r\rightarrow\infty)&=b_{00}+j\ \mathrm{log}(r),\\
    \beta_{2}(r\rightarrow\infty)&=b'_{00}+\frac{2k\omega}{k^2-\omega^2}j\ \mathrm{log}(r),\\
\end{aligned}
\end{split}
\end{equation}
where we have kept the terms up to the first sub-leading order.

The boundary current $J^1$ at leading order is
\begin{equation}
\begin{split}
\begin{aligned}
    J^{1}&=J^{\mu\nu}\frac{\delta b_{\mu\nu}}{\delta\beta_{1}}\\
    &\sim\frac{j}{\gamma^2(k^2-\omega^2)}=\frac{r\beta_{1}'}{\gamma^2(k^2-\omega^2)}.\\
\end{aligned}
\end{split}
\end{equation}
With $J^1$ we rewrite the boundary solution in eq. \eqref{bdry_solution_massless_2_form_5d} as
\begin{equation}\label{bdry_solution_massless_2_form_5d(1)}
    \beta_{1}(r\rightarrow\infty) = b_{00}+\gamma^2(k^2-\omega^2)J^1 \mathrm{log}(r).
\end{equation}
With the double trace deformation in eq. \eqref{double_trace_deform}, we obtain a shift term identical to the one for $\mathrm{AdS}_{4}$,
\begin{equation}
\begin{split}
\begin{aligned}
    \frac{\delta S_{ct}}{\delta J^1}&= \frac{\gamma^2}{2\kappa}\frac{\delta J_{\mu\nu}J^{\mu\nu}}{\delta J^1}\\
     &\sim -\frac{j}{\kappa}.\\
\end{aligned}
\end{split}
\end{equation}
At some cutoff $r_{\Lambda}$, this shift term and the boundary solution in eq. \eqref{bdry_solution_massless_2_form_5d(1)} yield
\begin{equation}
\begin{split}
\begin{aligned}
     \beta_1(r_{\Lambda})&=b_{00}+\gamma^2(k^2-\omega^2)J^1 \mathrm{log}(r_{\Lambda})- \frac{\gamma^2(k^2-\omega^2)J^1}{\kappa}\\
     &\equiv b_{00} - \frac{\gamma^2(k^2-\omega^2)J^1}{\kappa},\\
\end{aligned}
\end{split}
\end{equation}
where in the second line we have redefined $b_{00}$ as the source.
Finally, we evaluate the correlator of $J^1$:
\begin{equation}
\begin{split}
\begin{aligned}
     G^R_{J^1,J^1}(\omega,k) &= \frac{J^1}{b_{00}}\\
     &=\frac{J^1}{\beta_1(r_{\Lambda})+\frac{\gamma^2(k^2-\omega^2)J^1}{\kappa}}\\
     &=\frac{1}{\gamma^2(k^2-\omega^2)[\frac{\beta_1(r_{\Lambda})}{r_{\Lambda}\beta_1'(r_{\Lambda})}+\frac{1}{\kappa}]}.\\
\end{aligned}
\end{split}
\end{equation}

In $\mathrm{AdS}_{5}$, we expect the correlator to be dependent only on the Landau pole $r_{*}=r_{\Lambda}e^{\frac{1}{\kappa}}$, an invariant through the RG flow \cite{Hofman:2017vwr}. This could be achieved by a possible combination of $r_{\Lambda}$ and $\kappa$ if one found an explicit formula for the correlator.
In addition, if we expand $\beta_1(r_{\Lambda})$ as the infinite series in the form of the near-horizon expansion, we can again draw the conclusion that $G^R_{J^1,J^1}$ has a series of IR pole-skipping points, the first of which is exactly $(\omega,k)$ in eq. \eqref{ps_massless_2_form}, unaltered by the shift term from the double-trace deformation.

\subsection{Massive 2-form}\label{4.2}

The 2-form action added with a quadratic term of a non-zero mass $m$ is
\begin{equation}
    S[A]=-\frac{1}{6\gamma^2}\int\mathrm{d}^{d+2} x \sqrt{-g}Z(\phi)(H^2+3 m^2 B^2).
\end{equation}
The corresponding EOM are
\begin{equation}\label{EOM_massive_2_form}
    \partial_M(\sqrt{-g}Z(\phi)H^{MNP})-\sqrt{-g}Z(\phi)m^2 B^{NP} = 0.
\end{equation}
along with constraint equation
\begin{equation}
    \partial_M(\sqrt{-g}Z(\phi) B^{MN}) = 0.
\end{equation}

Similar to the massless case, the boundary current operator $J$ is now evaluated as
\begin{equation}\label{current_massive_2_form}
    J^{\mu\nu} = -\frac{1}{\gamma^2} \sqrt{-g}Z(\phi) (H^{r\mu\nu} + m^2 B^{\mu\nu}).
\end{equation}
With the same ansatz of near-horizon expansion as eq. \eqref{horizon_ansatz_2_form}, we find the first special points for the longitudinal mode containing $b_{v\mathbf{z}},\;b_{r\mathbf{z}},\;b_{x\mathbf{z}}$ components ($\mathbf{z}$ denotes an arbitrary spatial coordinate except $x$)
\begin{equation}\label{ps_massive_2_form}
\begin{split}
\begin{aligned}
    \omega &= -2i\pi T,\\
    k^2 &= -\left( m^2 h(r_0)+ \pi T h'(r_0) \pm \pi T\sqrt{\left[(d-3)h'(r_0)+2h(r_0) \frac{Z'(r_0)}{Z(r_0)}\right]^2+\frac{4m^2}{\pi T}h(r_0)h'(r_0)}\right).\\
\end{aligned}
\end{split}
\end{equation}
In $m\rightarrow 0$ limit, one of the ($k^2$) reduces to the special points of the transverse mode of the massless 1-form, cf. eq. \eqref{ps_massless_1_form}, and the other reduces to the special points of the longitudinal mode of the massless 2-form, cf. eq. \eqref{ps_massless_2_form}. Again we find that the appearance of a non-zero mass breaks the gauge symmetry and yields extra pole-skipping points. We will elaborate on this issue in the discussion section. 
\newline Now we demonstrate that the mass-dependent boundary divergence of 2-form in the bulk does not affect the pole-skipping properties of the boundary correlators either. For brevity, performing the same double-trace method as done in the massless model, we renormalize the boundary behavior and evaluate the boundary correlators, by assuming asymptotic $\mathrm{AdS}_{4}$, for instance, to be the background spacetime. 
We expand the field components on the boundary as
\begin{equation}\label{bdry_ansatz_2_form(1)}
    b_{\mu\nu}(r\rightarrow\infty)=r^{\lambda_{\mu\nu}}(b_{\mu\nu,00}+\frac{b_{\mu\nu,10}+b_{\mu\nu,11}\mathrm{log}(r)}{r}+\cdots).\\
\end{equation}
The boundary solution of $b_{v\mathbf{z}}$ is easily found to be 
\begin{equation}
    b_{v\mathbf{z}}(r\rightarrow\infty) = jr^{\lambda},\; \lambda=\frac{1}{2}(1+\sqrt{1+4m^2}),
\end{equation}
and $b_{x\mathbf{z}}$ is of the same order. 
Then from eq. \eqref{current_massive_2_form}, the boundary current $J^{v\mathbf{z}}$ is at leading order
\begin{equation}
    J^{v\mathbf{z}}= \frac{j\lambda r^{\lambda-1}}{\gamma^2}.\\
\end{equation}
With the double trace deformation in eq. \eqref{double_trace_deform}, we obtain a shift term
\begin{equation}
    \frac{\delta S_{ct}}{\delta J^{v\mathbf{z}}} = \frac{\gamma^2}{2\kappa}\frac{\delta J_{\mu\nu}J^{\mu\nu}}{\delta J^{v\mathbf{z}}} = -\frac{\gamma^2 J^{v\mathbf{z}}}{\kappa}.
\end{equation}
Therefore we renormalize the boundary solution at some cutoff $r_{\Lambda}$ to be
\begin{equation}
     b_{v\mathbf{z}}(r_{\Lambda}) \equiv b_{00} - \frac{\gamma^2}{\kappa}J^{v\mathbf{z}},\\
\end{equation}
where $b_{00}$ is the redefined source. Finally, we evaluate the correlator of $J^{v\mathbf{z}}$:
\begin{equation}
\begin{split}
\begin{aligned}
     G^R_{J^{v\mathbf{z}},J^{v\mathbf{z}}}(\omega,k) &= \frac{J^{v\mathbf{z}}}{b_{00}}\\
     &=\frac{J^{v\mathbf{z}}}{b_{v\mathbf{z}}(r_{\Lambda})+\frac{\gamma^2 J^{v\mathbf{z}}}{\kappa}}\\
     &=\frac{1}{\gamma^2[\frac{b_{v\mathbf{z}}(r_{\Lambda})}{b_{v\mathbf{z}}'(r_{\Lambda})}+\frac{1}{\kappa}]}.\\
\end{aligned}
\end{split}
\end{equation}
Note that this abstract form looks identical to the massless correlator in eq. \eqref{GF_massless_2_form_4d}, only different in an overall prefactor. But its explicit form depends on the bulk field EOM now with an additional mass term. Insert the near-horizon expansion of $b_{v\mathbf{z}}(r_{\Lambda})$ in eq. \eqref{horizon_ansatz_2_form} and we shall verify that the corresponding correlator has a series of IR pole-skipping points, the first of which is $(\omega,k)$ in eq. \eqref{ps_massive_2_form}, unaltered by the shift term from the double-trace deformation. As an aside, one can check that in higher spacetime dimensions, this statement also holds.

\section{\textit{p}-form}\label{5}

Now we generalize the discussion of the previous two sections to $p$-form to probe the pole-skipping property with near-horizon analysis. In subsection \ref{5.1}, we study the massless $U(1)$-gauge case; in subsection \ref{5.2}, we briefly talk about the massive configuration. 

\subsection{Massless \textit{p}-form}\label{5.1}

Generically, the action term of a minimally-coupled $p$-form field $P$ can be written as
\begin{equation}
    S[P]=-\frac{1}{2(p+1)}\int\mathrm{d}^{d+2} x \sqrt{-g}Z(\phi)(\mathrm{d}P)^2,
\end{equation}
where $\mathrm{d}P$ is the $(p+1)$-form field strength, with components
\begin{equation}\label{dP}
\begin{split}
\begin{aligned}
    \mathrm{d}P_{M_0\cdots M_p} &= (p+1)\partial_{[M_0}P_{M_1\cdots M_p]} \\
    &= \frac{1}{p!}\mathcal{P}(\partial_{M_0}P_{M_1\cdots M_p}) \\
    &= \displaystyle\sum_{j=0} ^{p} (-1)^j \partial_{M_j} P_{M_0\cdots M_{j-1}M_{j+1} \cdots M_p},
\end{aligned}
\end{split}
\end{equation}
where in the second line $\mathcal{P}$ denotes the alternating sum over all permutations of the indices. The theory has $U(1)$ gauge symmetry, i.e., $P \rightarrow P + \mathrm{d} \Lambda$ for arbitrary $(p-1)$-form $\Lambda$ whose exterior derivative obeys eq. \eqref{dP}. By taking the functional derivatives, we obtain EOM
\begin{equation}
    \partial_{M_0}(\sqrt{-g}Z(\phi)\mathrm{d}P^{M_0\cdots M_p}) = 0.
\end{equation}
We still assume that the wave vector of the field lies in $x$ direction, and that the field has a plane-wave expansion, i.e., 
\begin{equation}\label{fourier_p_form}
    P(r,v,x)=\int \mathrm{d}^2k \; e^{-i\omega v+ikx}\mathsf{p}(r,\omega,k).
\end{equation}
As a natural generalization from 2-form, with the gauge condition $P_{r\cdots}=0$, there are generically three modes for $p$-form $P$, which we name as \footnote{The components of a higher form generically transform as tensors under $SO(d-1)$. The three modes are named hereby so as to be consistent with 2-form.} (1) null mode $\mathsf{p}_{vx\cdots}$, which is trivial, (2) longitudinal mode $\mathsf{p}_{v\mathbf{z_1}\cdots\mathbf{z_{p-1}}}, \mathsf{p}_{x\mathbf{z_1}\cdots\mathbf{z_{p-1}}}$, where $\mathbf{z}$ denotes any spatial coordinate except $x$, (3) transverse mode $\mathsf{p}_{\mathbf{z_1}\cdots\mathbf{z_p}}$, which exists only in dimension $d\geq p+1$. Note that such a classification of modes only holds for the form number $p\geq 2$. Then one can check the EOM of the longitudinal mode and the transverse mode
\begin{equation}\label{EOM_massless_p_form}
\begin{split}
\begin{aligned}
    0 &= k^2 \mathsf{p}_{v\mathbf{z_1}\cdots\mathbf{z_{p-1}}} + k\omega \mathsf{p}_{x\mathbf{z_1}\cdots\mathbf{z_{p-1}}} + i\omega h \mathsf{p}_{v\mathbf{z_1}\cdots\mathbf{z_{p-1}}}' + ik r^2 f \mathsf{p}_{x\mathbf{z_1}\cdots\mathbf{z_{p-1}}}',\\
    0 &= ikZ \mathsf{p}_{x\mathbf{z_1}\cdots\mathbf{z_{p-1}}}' + ((d/2-(p-1))Z h'+h Z')\mathsf{p}_{v\mathbf{z_1}\cdots\mathbf{z_{p-1}}}' + hZ \mathsf{p}_{v\mathbf{z_1}\cdots\mathbf{z_{p-1}}}'',\\
    0 &= (Z(2k^2+i(d-2p)\omega h')+2i\omega hZ')\mathsf{p}_{\mathbf{z_1}\cdots\mathbf{z_p}} - 2r^2 fhZ \mathsf{p}_{\mathbf{z_1}\cdots\mathbf{z_p}}''\\
      &+ ((2p-d)r^2 fZh'-2h(Z(-2i\omega+2rf+r^2 f')+r^2 fZ')) \mathsf{p}_{\mathbf{z_1}\cdots\mathbf{z_p}}'.\\
\end{aligned}
\end{split}
\end{equation}
Imposing the ansatz of near-horizon expansion for $\mathsf{p}_{\mu_{1}\cdots\mu_{p}}$
\begin{equation}\label{horizon_ansatz_p_form}
     \mathsf{p}_{\mu_{1}\cdots\mu_{p}}(r)=(r-r_0)^{\lambda_{\mu_{1}\cdots\mu_{p}}}(\mathsf{p}_{\mu_{1}\cdots\mu_{p}}^{(0)}+\mathsf{p}_{\mu_{1}\cdots\mu_{p}}^{(1)}(r-r_0)+\mathsf{p}_{\mu_{1}\cdots\mu_{p}}^{(2)}(r-r_0)^2+\cdots),
\end{equation}
or analogously for the gauge invariants, we find the first-order pole-skipping points for the boundary correlators of the longitudinal and transverse modes at
\begin{equation}\label{ps_massless_p_form}
\begin{split}
\begin{aligned}
    \omega&=-2i\pi T,\ \ k^2=\pi T((d-2p)h'+2hZ'/Z),\quad \text{for longitudinal mode},\\
    \omega&=-2i\pi T,\ \ k^2=-\pi T((d-2p)h'+2hZ'/Z),\ \text{for transverse mode},\\
\end{aligned}
\end{split}
\end{equation}
where $h$ and $Z$ are evaluated at $r=r_0$. Similar to the case of 1-form, the $p$-form longitudinal mode has an extra zeroth order pole-skipping point at $\omega=k=0$.

\subsection{Massive \textit{p}-form}\label{5.2}

In analogy, we briefly present the computation of bulk $p$-form with a non-zero mass. The action term is 
\begin{equation}
    S[P]=-\frac{1}{2(p+1)}\int\mathrm{d}^{d+2} x \sqrt{-g}Z(\phi)((\mathrm{d}P)^2 + (p+1)m^2 P^2).
\end{equation}
The corresponding field EOM are
\begin{equation}
    \partial_{M_0}(\sqrt{-g}Z(\phi)\mathrm{d}P^{M_0\cdots M_p})-\sqrt{-g}Z(\phi)m^2 P^{M_1\cdots M_p} = 0.
\end{equation}
Following the simplification procedure of 1-form and 2-form and assuming the plane-wave expansion in eq. \eqref{fourier_p_form} and the near-horizon expansion of the field components in eq. \eqref{horizon_ansatz_p_form}, we find the first pole-skipping points for the longitudinal mode at
\begin{small}
\begin{equation}\label{ps_massive_p_form}
\begin{split}
\begin{aligned}
    \omega &= -2i\pi T,\\
    k^2 &= -\left( m^2 h(r_0)+ \pi T h'(r_0) \pm \pi T\sqrt{\left[(d+1-2p)h'(r_0)+2h(r_0) \frac{Z'(r_0)}{Z(r_0)}\right]^2+\frac{4m^2}{\pi T}h(r_0)h'(r_0)}\right).\\
\end{aligned}
\end{split}
\end{equation}
\end{small}
In $m\rightarrow 0$ limit, one of the ($k^2$) reduces to the special points of the transverse mode of the massless ($p-1$)-form, and the other reduces to the special points of the longitudinal mode of the massless $p$-form, cf. eq. \eqref{ps_massless_p_form}. Again we find that the appearance of a non-zero mass breaks the gauge symmetry and yields extra pole-skipping points. 

\section{Discussion and conclusion}\label{6}

In this section, we briefly summarize the main results aforesaid, and clarify some facts about the pole-skipping properties of the $U(1)$ gauge models and those symmetry-broken massive configurations separately.

We have found the pole-skipping points of $U(1)$-gauged boundary correlators with near-horizon analysis in the bulk spacetime. As explained in this work as well as in \cite{Blake:2019otz}, at these points on the $(\omega,k)$-plane, the two independent solutions degenerate, or more precisely, the extra d.o.f. in the coefficients of the near-horizon series expansion render the solution unphysical. Specifically, to study the full dynamics in asymptotic $\mathrm{AdS}_{d+2}$, we have considered the $U(1)$ field configured as 1-form with $d\geq 2$ and higher $p$-form with $d\geq p+1$. The first-order pole-skipping points of the boundary correlators are located at 
\begin{equation}\label{ps_massless_forms}
\begin{split}
\begin{aligned}
    \omega&=-2i\pi T,\ \ k^2=\pi T((d-2p)h'+2hZ'/Z),  &\text{for }\textit{p}\text{-form longitudinal mode},\\
    \omega&=-2i\pi T,\ \ k^2=-\pi T((d-2p)h'+2hZ'/Z), &\text{for }\textit{p}\text{-form transverse mode},\\
\end{aligned}
\end{split}
\end{equation}
considering the transformation properties of the modes under corresponding symmetry groups of the background spatial sub-manifold. The $p$-form longitudinal mode has an extra zeroth order pole-skipping point at $\omega=k=0$. This is a generalization of the results in \cite{Blake:2019otz, Natsuume:2019xcy}. Now we seek to find a relation between the pole-skipping properties of $U(1)$ correlators and electromagnetic duality. Recall the proportionality of the boundary values of the dual fields in eq. \eqref{dual_prop}
\begin{equation}
    \Tilde{\beta}^{(d-p)} \propto \ast \mathrm{d}\alpha^{(p+1)},\quad \mathrm{d}\Tilde{\alpha}^{(d+1-p)} \propto \ast \beta^{(p)},
\end{equation}
and we can establish with our plane-wave expansion in eq. \eqref{fourier_p_form} a pair of dual relations of their components evaluated on the boundary:
\begin{equation}\label{dual_modes}
    \mathsf{p}_{\mathbf{z_1}\cdots\mathbf{z_{p}}} \Longleftrightarrow \mathsf{q}_{v\mathbf{z_{p+1}}\cdots\mathbf{z_{d-1}}},\quad 
    \mathsf{p}_{v\mathbf{z_1}\cdots\mathbf{z_{p-1}}} \Longleftrightarrow \mathsf{q}_{\mathbf{z_p}\cdots\mathbf{z_{d-1}}}.
\end{equation}
These relations prompt us to investigate the pole-skipping points of the corresponding modes of $P$ and its dual $Q$. Recall the points for the scalar (0-form) correlator in eq. \eqref{ps_massive_scalar}, which take the values in zero-mass limit
\begin{equation}\label{ps_massless_scalar}
    \omega = -2i\pi T,\quad k^2 = -m^2 h - \pi Tdh' \stackrel{m\rightarrow 0}{\longrightarrow} -\pi Tdh'.\\
\end{equation}
With some simple calculations, we discover that the pole-skipping points of the dual modes are exactly identical if the dilatonic coupling is ignored, by using eq. \eqref{ps_massless_forms}, \eqref{dual_modes}, and \eqref{ps_massless_scalar}. For instance,
\begin{equation}
\begin{split}
\begin{aligned}
    \text{in asymptotic AdS}_4:\;& k^2(b_{longi}) = k^2(\phi),\\
    \text{in asymptotic AdS}_5:\;& k^2(c_{longi}) = k^2(\phi),\\
                                 & k^2(b_{longi}) = k^2(a_{trans}),\;k^2(b_{trans}) = k^2(a_{longi}),\\
    \text{in asymptotic AdS}_{d+2}:\;& k^2(d_{longi}) = k^2(\phi),\\
     & k^2(\mathsf{p}_{longi}) = k^2(\mathsf{q}_{trans}),\;k^2(\mathsf{p}_{trans}) = k^2(\mathsf{q}_{longi}),\\
\end{aligned}
\end{split}
\end{equation}
where we have denoted the correlators of the scalar, 1-form, 2-form, 3-form, $d$-form, and $(p\geq 2)$-form by $\phi,\;a,\;b,\;c,\;d,\;\mathsf{p}$, respectively, and denoted the longitudinal and transverse modes by the shorthand of their first five letters. Therefore, we declare a trans-mode identification of pole-skipping, which we consider as a near-horizon verification of electromagnetic duality in asymptotic AdS geometry.

Similarly, we have probed the generalized configuration by assuming a non-zero mass for the bulk fields. For simplicity, we have focused on the longitudinal, whose first-order pole-skipping points are located at
\begin{small}
\begin{equation}\label{ps_massive_forms}
\begin{split}
\begin{aligned}
    \omega &= -2i\pi T,\\
    k^2 &= -\left( m^2 h(r_0)+ \pi T h'(r_0) \pm \pi T\sqrt{\left[(d+1-2p)h'(r_0)+2h(r_0) \frac{Z'(r_0)}{Z(r_0)}\right]^2+\frac{4m^2}{\pi T}h(r_0)h'(r_0)}\right).\\
\end{aligned}
\end{split}
\end{equation}
\end{small}
In $m\rightarrow 0$ limit, one of the ($k^2$) reduces to the special points of the transverse mode of the massless ($p-1$)-form, and the other reduces to the special points of the longitudinal mode of the massless $p$-form, cf. eq. \eqref{ps_massless_p_form}. We find that the appearance of a non-zero mass breaks the gauge symmetry and yields extra pole-skipping points. Note that the $p$-form longitudinal mode has an extra zeroth order pole-skipping point at $\omega=0,\;k=-m^2 h(r_0)$. In $m\rightarrow 0$ limit, this point reduces to ($\omega$=k=0), i.e. the zeroth order point of the longitudinal mode of the massless configurations. 

As argued in \cite{Struyve:2011nz}, in gauge field theories the ground state would be degenerate, with equivalent states connected by the gauge symmetry. In our context, the explicitly broken gauge symmetry gives rise to new physics, say, prospectively in the pole-skipping phenomenon which could have a correspondence to the non-degeneracy of the boundary ground state in chaotic systems.

Furthermore, we have reviewed with the example of 2-form that the boundary divergence can be fixed by holographic renormalization, or more precisely, by means of the double-trace deformation in the view of the boundary CFT. Therefore, with the choice of mixed boundary condition, the redefined source and the double-trace coupling term would lead to well-defined holographic correlators. Then we have demonstrated that the coupling $\kappa$ does not make a difference in the pole-skipping properties considering the IR physics, i.e., the near-horizon boundary condition. We understand the meaning of IR in the sense that:

(1) By computing the correlators, one can tell directly where their pole-skipping points lie. Such kind of computation is practically difficult both in CFT and by GKPW relation (which is simplified as the ratio of the response and the source in the linear response system). Compared to quantized boundary CFT, the gravitational computation in the bulk is IR. (2) In another sense, the renormalization of the boundary CFT corresponds to the radial cutoff in the bulk, that is, the flow downward to lower energy scale is translated to be the radial cutoff deeper in the bulk \cite{Faulkner:2010jy, deBoer:1999tgo, Balasubramanian:1999jd}. Therefore, the pole-skipping points of the boundary correlators only depend on IR bulk physics.

As a complement, we have introduced in eq. \eqref{bdry_p_form} from \cite{DeWolfe:2020uzb} that with the gauge condition $P_{r\cdots}=0$, in $\mathrm{AdS}_{d+2}$ a generic propagating $p$-form field $P$ behaves on the boundary as \footnote{In \cite{DeWolfe:2020uzb}, the boundary spacetime is assumed to be $\mathrm{AdS}_{d+1}$. Please note this number difference from our setup for a close examination.}
\begin{equation}
    P(r\rightarrow\infty) = \alpha + \beta \ r^{2p-d-1} + \cdots,
\end{equation}
where $\alpha$ and $\beta$ are constants of $r$. Exceptionally, a logarithmic term would appear when $d$ is odd and $p=(d-1)/2$ \cite{DeWolfe:2020uzb}, exemplified by the boundary solution of 2-form, the electromagnetic dual, for $d=3$ as we discussed in section \ref{4}. Therefore, one could adopt a similar prescription of renormalization to fix the possibly existing boundary divergence and thus to examine the special points obtained from near-horizon analysis in eq. \eqref{ps_massless_p_form} and eq. \eqref{ps_massive_p_form}.

In conclusion, we have studied the pole-skipping properties of $U(1)$-gauged holographic correlators of form currents and discovered a trans-mode equivalence in correspondence with electromagnetic duality. We have discussed the generalization with explicit symmetry breaking, indicative of more interesting chaotic physics to be detected. We have also substantiated our near-horizon computation by holographically renormalizing the boundary solutions and demonstrated that the pole-skipping properties are only determined by the IR physics, consistent with the previous models without divergence on the boundary. These results motivate us to extend our studies to models with spontaneous symmetry breaking in future works, as an attempt to investigate the superconducting phase with holography.

\section*{Acknowledgement}

We thank Prof. Yan Liu and Prof. Ya-Wen Sun for their suggestions on the project and helpful guidance throughout the work.
We are grateful to Dr. Hyun-Sik Jeong for thoughtful consideration and helpful advice. We benefit immensely from the discussion with Yuan-Chun Jing, Dian-Dian Wang, Zi-Yue Wang, and Xin-Xiang Ju. This work was supported by the National Key R\&D Program of China (Grant No. 2018FYA0305800), National Natural Science Foundation of China (Grant No. 12035016), and the Strategic Priority Research Program of Chinese Academy of Sciences (Grant No. XDB28000000).

\paragraph{Open Access}

This article is distributed under the terms of the Creative Commons Attribution License (CC-BY 4.0), which permits any use, distribution, and reproduction in any medium, provided the original author(s) and source are credited.

\bibliography{biblio}
\bibliographystyle{JHEP}

\end{document}